# Electro-Reflectance Spectra of Blue Bronze

R.C. Rai[1], V.A. Bondarenko[1,2], and J.W. Brill[1a]
[1] Department of Physics and Astronomy, University of Kentucky, Lexington, KY 40506-0055, USA
[2] Institute of Semiconductor Physics, National Academy of Sciences of Ukraine, 45 Prospect Nauki, Kiev, 03028 Ukraine

**Abstract.**  We show that the infrared reflectance of the quasi-one dimensional charge-density-wave (CDW) conductor $K_{0.3}MoO_3$ (blue bronze) varies with position when a voltage greater than the CDW depinning threshold is applied.  The spatial dependence and spectra associated with these changes are generally as expected from the electro-transmission [B.M. Emerling, *et al*, Eur. Phys. J. B **16**, 295 (2000)], but there are differences which might be associated with changes in the CDW properties on the surface.  We have examined the electro-reflectance spectrum associated with CDW current investigation for light polarized parallel to the conducting chains for signs of expected current-induced intragap states, and conclude that the density of any such states is at least an order of magnitude less than expected.

[a]email: jwbrill@uky.edu

Quasi-one-dimensional conductors with sliding charge-density-waves (CDW's) exhibit a variety of unusual electronic properties, resulting from the deformation and motion of the CDW in an applied electric field. In general, an incommensurate CDW is pinned to the lattice by impurities. Small applied fields polarize the CDW, i.e. cause it to strain from its pinned configuration; above a depinning voltage, $V_T$, the CDW slides through the sample, carrying collective current [1].

A long-standing question has concerned the mechanism by which electrons enter and leave the sliding CDW at current contacts. X-ray [2] and transport measurements [3] have studied the spatial dependence of the CDW current and deformations, and have shown that current conversion involves additional CDW strains at the contacts above the more gradual, "bulk" CDW polarization strain; both bulk and contact strains have also been observed through position dependent changes in the infrared transmission [4]. The CDW strain is proportional to $\partial\varphi/\partial x$, where x is the coordinate in the high conductivity direction, and $\varphi$ the local CDW phase [1]. The contact strains are believed to drive CDW phase-slip through the growth and climb of "macroscopic" CDW dislocation loops [2,3,5]. Microscopically, these loops are thought to form from the condensation of $2\pi$–phase-solitons, with energy levels near the edge of the CDW gap, on the individual chains [6]. The $2\pi$-solitons, corresponding to 2 electrons entering or leaving the CDW, in turn are believed to form (within $\approx$ 1 ps) from the condensation of 1-electron midgap $\pi$-solitons [6]. However, such current-induced intragap soliton/dislocation states have not been observed, although related states arising as "growth defects" upon cooling or doping may have been seen spectroscopically [7,8].

We recently reported on the results of a search for such states using electromodulated infrared transmission spectroscopy [9] in the quasi-one-dimensional conductor $K_{0.3}MoO_3$ (blue bronze [1,10]) below its CDW transition temperature, $T_c$ = 180 K. As mentioned above, the infrared transmission ($\tau$) varies spatially in response to an applied voltage. Near the center of the sample, the relative change in transmission ($\Delta\tau/\tau$) varies approximately linearly with position (with the transmission increasing on the positive side of the sample), reflecting the bulk polarization of the CDW, but larger changes are observed near (~ 0.1 mm) the current contacts [4], as shown in the lower inset to Figure 1. We suggested [11] that the variation in $\Delta\tau/\tau$ largely reflects changes in intraband absorption by thermally excited quasiparticles (mostly electrons in blue bronze) whose local density varies to screen CDW deformations so that $\Delta\tau/\tau \propto \partial\varphi/\partial x$, although pronounced spectral changes in $\Delta\tau/\tau$ associated with small changes of phonon frequencies, lifetimes, and oscillator strengths caused by the strained CDW were also observed [9], as shown in the inset to Figure 2. We compared the bulk and contact electromodulated transmission spectra to search for new absorption features in the latter that could be associated with intragap electronic states caused by current injection into the CDW. No such absorption features were observed [9], giving an upper limit to [$n\sigma/\Gamma$], where n is the soliton density, $\sigma$ its optical cross-section, and $\Gamma$ its linewidth:

$$n\sigma/\Gamma \sim \varepsilon_1^{1/2} \delta(\Delta\tau/\tau) / \gamma d < 4 \times 10^{-10} (\text{Å cm}^{-1})^{-1}, \qquad (1)$$

where $\delta(\Delta\tau/\tau)$ was the difference in the (normalized) contact and bulk electromodulated transmission signals, d ~ 3.5 μm the thickness of the sample, $\gamma$ the spectral width of the source (~ 10 cm$^{-1}$), and $\varepsilon_1 \sim 4$ [10,12] is the real part of the background dielectric constant [13]. Using the expected values of $\Gamma \sim k_B T_c$ [6] and $\sigma \sim \xi_{//}\xi_\perp \sim 100$ Å$^2$, its geometric cross-section, where $\xi_{//}$ and $\xi_\perp$ are the longitudinal and transverse CDW amplitude coherence lengths [14], this implies that midgap states occur on less than 4% of the chains, whereas we expect all chains to have such states for large CDW currents (see below). However, because of the strong absorption of light polarized parallel to the conducting chains, the transmission spectra could only be measured for transversely polarized light, and it was suggested [9] that the optical cross-section might be much smaller than the geometric cross-section for this polarization. These results motivated a search for voltage induced changes in the infrared reflectivity and measurements of its spectra, especially for parallel polarized light. The results of these measurements are reported here.

The 2π-soliton states are believed to have energies within ~ $k_B T_c$ ~ 125 cm$^{-1}$ of the edge of the CDW gap, 2Δ [6]. Unfortunately, the value of the gap has not been unambiguously determined. Previous estimates of 2Δ from different measurements have varied from 800 cm$^{-1}$ to 1400 cm$^{-1}$ [10,12,15,16], perhaps reflecting the effects of quantum and thermal fluctuations in broadening the absorption edge [16]. In Reference [9], we present evidence that 2Δ ~ 1150 cm$^{-1}$ (close to the early optical estimates [10,12]) is an appropriate choice for the electro-optic response.

Smooth sample surfaces were prepared by cleaving as grown single crystals of blue bronze with sticky tape and etching them with dilute ammonium hydroxide. Gold films were evaporated on the ends of the crystals for current contacts, separated by ~ 1mm (see Figure 1, upper inset). Electrical and thermal contacts to the sample were made by attaching its ends with silver paint to a sapphire substrate in a liquid nitrogen cryostat. All measurements were made with the sapphire substrate at 77 K, although the sample may have been a few degrees warmer than this. Four tunable infrared (PbSnSe) diode lasers, cooled with a helium refrigerator and covering the range 400 cm$^{-1}$ to 1200 cm$^{-1}$ with a typical power of 0.5 mW, were used as light sources. The lasers were not mode selected and typically there were several modes spread over $\gamma \sim \pm 7$ cm$^{-1}$. An IR microscope (Nicolet Continuum) was used to vary the position of the illuminated light spot (typically 50 x 50 μm$^2$) on the sample.

Square-wave voltages were applied to the sample, and changes in the reflected intensity (in phase with the applied voltage) were measured with a lock-in amplifier operating at the square wave frequency, which was kept high enough (253 Hz) so that signals due to modulations of the sample's temperature were negligible. (We note that there was often also a small electro-reflectance response in quadrature with the applied voltage, reflecting the time delay of the CDW polarization [11] and residual thermal effects.) Simultaneously, the intensity of the reflected light, chopped at a different frequency, was measured with a second lock-in amplifier operating at the chopping frequency. Although neither the reflectance (R) nor its change with voltage (ΔR) was measured precisely by itself, the ratio of the two lock-in signals provides a precise measurement of the relative change, ΔR/R. (The gold current contacts on the samples provided references for absolute reflectivity spectra, but these measurements were "rough" because of imperfect films and surfaces, possible diffraction effects, and lack of reproducibility of the lasers.) The vibration of the lasers in the helium refrigerator typically limited our sensitivity in ΔR/R to ~ ± 5 x 10$^{-6}$.

Although our ultimate goal was to search for differences in the parallel polarized contact and bulk electro-reflectance spectra, we first examined the spectrum and spatial dependence for the transverse polarization, to verify that the reflectivity behaved as expected from the transmission. Figure 1a shows the spatial dependence of $\Delta R/R$ (at 1050 cm$^{-1}$) for transversely polarized light for sample 1 (with a threshold voltage $V_T = 15$ mV). Shown are the variations for bipolar square waves of amplitudes 40 mV and 16 mV (with $\Delta R \equiv [R(+V) - R(-V)]$) and a 40 mV positive unipolar square wave (with $\Delta R \equiv [R(+V) - R(0)]$).

These results, although expected in view of our transmission experiments, are the first observation of position dependent changes in the reflectivity associated with CDW depinning. (We note, however, that spatially uniform changes in the far infrared reflectivity, including resonance-like features associated with CDW motion through the lattice, have been reported [17].) The 40 mV spatial dependences are basically similar to what was observed in transmission experiments [4,9]. In particular, for the unipolar square-wave the CDW stays pinned and the polarization does not relax when the voltage is zero, so that only electromodulated signals due to the contact strains (associated with CDW motion) are observed, whereas for the bipolar square wave both contact and bulk signals are observed. One difference between these reflectivity spatial profiles and our previous transmission profiles [4], as shown in the lower inset to the Figure, is that decreases in $|\Delta R/R|$ are observed adjacent to the contacts. A similar spatial profile is observed for the parallel polarization, shown in Figure 1b. As shown by the 16 mV profile (open triangles), this drop near the contacts is much deeper at smaller voltages, so that the electro-reflectance signal actually becomes inverted. No corresponding effect was observed for the transmission [4], suggesting that the CDW polarization is different on the surface (probed by reflectance) than throughout the volume (probed by transmission), e.g. because of inhomogeneous current flow (e.g. current is injected from the gold contacts on the surface), surface reconstruction of the CDW, and/or surface pinning of the CDW. Future experiments on samples thin enough to measure both reflectance and transmission will allow us to more systematically study the differences between the transmission and reflectance spatial profiles as functions of voltage, square-wave frequency, and wavelength.

Figure 2b shows the electromodulated reflectivity spectrum for the transverse polarization (at V = ± 80 mV, 37 μm from the contact). The reflectivity spectrum, similar to that obtained by others [10,12], is shown in Figure 2a (plotted logarithmically to facilitate analysis of the electro-reflectance). As expected from the electromodulated transmission spectrum, there are sharp features in the electro-reflectance associated with phonon lines, labeled with letters (A-J) as in reference [9]. In fact, the phonon features dominate the electro-reflectance spectrum much more than for the electro-transmission spectrum [9], for which there was also a large, positive broadband signal, as shown in the inset to the Figure. However, whereas changes in phonon linewidths ($\Gamma_j$), frequencies ($\nu_j$), and oscillator strengths ($A_j$) lead to very distinct changes (2$^{nd}$-derivative, 1$^{st}$-derivative, and 0$^{th}$-derivative lineshapes, respectively) in the electro-transmission spectrum [9], as seen in the inset, all three result in more complex and less distinct changes in the electro-reflectance spectrum, complicated further by the breadth and overlap of the Fano reflectance lines. This makes it especially difficult to determine the

*signs* of the changes in phonon properties, but the magnitudes of the changes can be roughly determined from

$$|\Delta\nu_j| \sim |\Delta\Gamma_j| \sim \Gamma_j |\Delta A_j/A_j| \sim \Gamma_j [\Delta R/R] / \delta \ln R, \quad (2)$$

where $\delta \ln R$ is the magnitude of the drop in lnR for that phonon. For all the phonons (except for F), we find $|\Delta\nu_j|$ and/or $|\Delta\Gamma_j|$ to be between ~ 0.01 and 0.04 cm$^{-1}$ (or for phonon A [9], a change in oscillator strength ~ 0.1%), similar to what was deduced from the electro-transmission [9]. However, there now appears to be a very large change ($|\Delta\nu_j|$ and/or $|\Delta\Gamma_j| \sim 0.1$ cm$^{-1}$) for F, for which no change could be observed in the transmission [9]. Whether these differences are sample (or temperature) dependent effects, artifacts of the analysis, or indications that phonon changes are different near the surface must be addressed by comparisons of the electro-transmission and electro-reflectance on the same samples with a more quantitative modeling of the spectra.

Figure 3a shows the parallel reflectance spectrum, similar to published results [10,12]; we denote 6 phonon modes with letters K-P. The bulk and contact electro-reflectance spectra, taken at the locations shown by the arrows in Figure 1b, are shown in Figure 3b, with the contact spectrum multiplied by a normalization factor $N = 1.9 \pm 0.3$ for direct comparison. (The value of *N* varied slightly for spectra taken with different lasers, possibly because the positions of the images of the microscope aperture varied slightly for different laser positions.) Very similar electro-reflectance spectra were obtained for a second sample. Even more than for the transverse polarization, the phonon features dominate the spectrum. While, as for the transverse polarization, it is difficult to associate the features in the electro-reflectance spectrum unambiguously with particular phonon changes, the electro-reflectance features can generally be associated with frequency or linewidth changes varying from ~ 0.01 cm$^{-1}$ to 0.2 cm$^{-1}$ or oscillator strength changes varying from < 0.1% to > 1%. The largest changes are associated with phonon O; however, they were calculated with the (questionable) assumption that the large peak in $\Delta R/R$ at 695 cm$^{-1}$ is associated with the tiny reflectance peak at 705 cm$^{-1}$ [12] rather than the much larger Fano resonance at 660 cm$^{-1}$.

Away from the phonon lines, e.g. at frequencies above 1000 cm$^{-1}$, the electro-reflectance signal has opposite signs for the two polarizations, and the magnitude for the transverse polarization is an order of magnitude larger than for the parallel. These differences can be understood in terms of the effects of changes in quasiparticle density on the reflectance. For example, in a Drude model of excited quasiparticles, a decrease in density (such as occurs near the positive contact) will decrease the imaginary part of the dielectric constant but increase the real part. For moderately anisotropic quasiparticle masses and moderate degrees of damping, this can result in an increase in the transverse reflectance and a much smaller decrease in the (higher) parallel reflectance, as observed. However, a more complete analysis of the reflectance spectra than has been attempted until now would be necessary to confirm this explanation.

As seen in the figure, the contact and bulk spectra are very similar, with no new absorption lines seen in the contact spectrum. The expected difference in the two spectra due to the presence of intragap solitons in the contact region is (compare Eqtn. 1):

$$\delta(\Delta R/R) \sim (\partial \ln R/\partial \alpha) \delta\alpha \sim N (\gamma/\nu) a(\varepsilon) [n\sigma/\Gamma], \quad (3)$$

where α is the absorptivity [d $\delta\alpha \sim \delta(\Delta\tau/\tau)$] and the function $a(\epsilon) \sim 2 \times 10^{-3}$ for the parallel complex dielectric constant near 1000 cm$^{-1}$ [10]. Over most of the spectrum, the contact and bulk spectra agree to within their scatter: $\delta(\Delta R/R) < 5 \times 10^{-6}$, so that, [n$\sigma$/Γ] < $10^{-9}$/( Å cm$^{-1}$). Obtaining a better limit would require having a more stable optical source.

The scatter in the data is somewhat larger near 900 cm$^{-1}$, and the apparent difference between the contact and bulk spectra shown in Figure 3b, $\delta(\Delta R/R) \sim 1.5 \times 10^{-5}$ in this region is not reproducible, so that [n$\sigma$/Γ] < $4 \times 10^{-9}$ /( Å cm$^{-1}$) for energies between 800 and 1000 cm$^{-1}$. The increased noise in this spectral region near the zero-crossing of $\Delta R/R$ may reflect noise in the phase of the CDW response [11] because, as mentioned above, there was a small quadrature signal.

Taking the expected values of $\Gamma \sim k_B T_c \sim 125$ cm$^{-1}$ and $\sigma \sim \xi_{//}\xi_{\perp} \sim 100$ Å$^2$, we find an upper limit for the soliton density of n < $1.3 \times 10^{15}$ cm$^{-3}$ for most of the spectrum, with the upper limit a few times greater than this near 900 cm$^{-1}$. While this limit is a few times greater than we obtained from the electro-transmission, our estimate for the cross-section is probably more appropriate for the parallel polarization, as discussed above. These intragap states are expected to be distributed throughout the phase-slip regions of length $\Lambda \sim 0.1$ mm near the contacts, i.e. where the unipolar electromodulated response is observed. Taking the area/conducting chain to be $\Omega \sim 100$ Å$^2$, our results imply that, for energies over most of the subgap spectrum, less than 15% of the conducting chains contain such an intragap state. For comparison, the change in electro-reflectance that would result from one soliton per conducting chain, assuming a Lorentzian peak in the dielectric constant (arbitrarily placed at 1100 cm$^{-1}$), is shown in Figure 3b.

In fact, the following simple model suggests that there should be roughly one soliton/chain. Two electrons, which form a 2π-soliton, enter each conducting chain in the time (*T*) it takes the CDW phase to advance 2π [6]. (This washboard, "narrow-band-noise" [1] period ~ 5 μs at V=50 mV in our samples.) These solitons condense into dislocation lines [6] which climb across the sample at a speed $v_{DIS}$, so that

$$n \sim (\Omega^{1/2}/ v_{DIS}T) / \Omega\Lambda. \quad (4)$$

If we *assume* that the dislocation is forced to climb by the advance of the CDW (with speed $v_{CDW}$), then

$$v_{DIS} / \Omega^{1/2} \sim v_{CDW}/ \lambda, = 1/T, \quad (5)$$

where λ is the CDW wavelength, and we find n ~ 1 / $\Lambda\Omega$. (This assumption presumably breaks down very close to threshold, where *T* becomes very long, and the soliton/dislocation motion becomes incoherent. However our observation of strong narrow-band-noise voltage oscillations at frequencies ~ 1 kHz for dc voltages only ~ 1 mV above threshold indicates that we are far from this limit. We also note that in their early work on CDW phase slip, Ong and Maki [18] assumed that the dislocations climb at the much faster, but sample dependent, speed, $v_{DIS} \sim v_{CDW} d/\lambda$, in which case the soliton density would be negligible.) Hence our results suggest that the soliton lifetime is an order of magnitude less than the narrow-band-noise period.

In conclusion, we present the first observation that the infrared reflectance of blue bronze varies with position when a voltage above the CDW depinning threshold is applied. While such variation is expected from the previous transmission experiments, the spatial dependence of the electro-reflectance differs from that of the electro-transmission, suggesting that the dynamics of pinning on the surface is different than in the bulk. As for the transverse polarization, we find that parallel polarized phonons are affected by CDW strain, and these changes dominate the electro-reflectance spectra. We compared the parallel polarized contact and bulk spectra to search for intragap states associated with CDW current injection and find an upper limit for [nσ/Γ] an order of magnitude lower than its expected value over most of the spectral range, suggesting that such states occur on less that 15% of the conducting chains.

We thank R.E. Thorne of Cornell University for providing samples. This research was supported by the United States National Science Foundation, Grant # DMR-0100572.

# References

1. G. Gruner, Rev. Mod. Phys. **60**, 1129 (1988).
2. H. Requardt, *et al*, Phys. Rev. Lett. **80**, 5631 (1998); S. Brazovskii, *et al*, Phys. Rev. B **61**, 10640 (2000).
3. T.L. Adelman, *et al*, Phys. Rev. B **53**, 1833 (1996); S.G. Lemay, *et al*, Phys. Rev. B **57**, 12781 (1998).
4. M. E. Itkis, B.M. Emerling, and J.W. Brill, Phys. Rev. B **52**, R11545 (1995).
5. J.C. Gill, Solid State Commun. **44**, 141 (1982); J. Phys. Condens. Matter **1**, 6649 (1989).
6. S. Brazovskii and S. Matveenko, J. Phys. I France **1**, 269, 1173 (1991).
7. G. Minton and J.W. Brill, Solid State Commun. **65**, 1069 (1988).
8. F.Ya. Nad' and M.E. Itkis, JETP Lett. **63**, 262 (1996).
9. B.M. Emerling, M.E. Itkis, and J.W. Brill, Eur. Phys. J. B **16**, 295 (2000).
10. G. Travaglini, *et al*, Solid State Commun. **37**, 599 (1981).
11. M.E. Itkis and J.W. Brill, Phys. Rev. Lett. **72**, 2049 (1994).
12. S. Jandl, *et al*, Phys. Rev. B **40**, 12487 (1989).
13. The factor of $\varepsilon_1^{1/2}$ was omitted in Reference [9].
14. G. Minton and J.W. Brill, Phys. Rev. B **45**, 8256 (1988).
15. L. Forro, *et al*, Phys. Rev. B **34**, 9047 (1986); D.C. Johnston, Phys. Rev. Lett. **52**, 2049 (1984); K. Nomura andK. Ichimura, J. Vac. Sci. Technol. **A8**, 504 (1990)
16. L. Degiorgi, et al, Phys. Rev. B **52**, 5603 (1995).
17. B. Gorshunov, *et al*, J. Phys IV France **12**, Pr9-81 (2002).
18. N.P. Ong and K. Maki, Phys. Rev. B **32**, 6582 (1985).

# Figure Captions

Fig. 1.  Spatial dependence of the relative change in reflectivity when applying 253 Hz square-wave voltages.  a) Light at 1050 cm$^{-1}$, transverse polarization.   Open circles: 40 mV, bipolar square-wave. Closed circles: 40 mV, positive unipolar square-wave.  Open triangles: 16 mV bipolar square-wave.  b) 715 cm$^{-1}$, parallel polarized light.  Open circle: 50 mV, bipolar square-wave.  Closed circles: 50 mV, positive unipolar square-wave.  The vertical arrows show the locations of the spectra shown in Figures 2 and 3.  The upper inset shows the configuration of the sample.  The lower inset shows the spatial dependence of the relative change in transmission for a sample of Ref. [4,9] measured with a broadband IR source: closed circles are for a unipolar voltage (3 $V_T$) and open circles are for a bipolar voltage ($\pm$ 3.7 $V_T$), with the line showing the "bulk" value of $\Delta\tau/\tau$.

Fig.2. a) Reflectivity spectrum for transverse polarization.  Phonon resonances are labeled A - J.    b) Electro-reflectance spectrum for transverse polarization, measured with 253 Hz, 80 mV bipolar square-wave at the location shown in Figure 1a.   The inset shows the transmission, $\tau$, and electro-transmission, $\Delta\tau/\tau$, spectra of a sample of Ref. [9] measured at threshold, so that $\Delta\tau \equiv \tau(+V_T) - \tau(-V_T)$.

Fig. 3.  a) Reflectivity spectrum for parallel polarization.  Phonon resonances are labeled K – P.  b) Electro-reflectance spectrum for parallel polarization, measured with 253 Hz, 50 mV square-waves at the locations shown in Figure 1b.  Open symbols: bipolar square-wave.  Closed symbols: unipolar square wave (multiplied by $N \sim 1.9$).  The curve is the calculated difference in the spectra (vertically offset for clarity) due to the presence of field-induced soliton excitations, with n = $10^{16}$/cm$^3$ ~ 1/chain, $\sigma$ = 100 A$^2$, $\Gamma$ = 125 cm$^{-1}$, and $\nu$ = 1100 cm$^{-1}$.

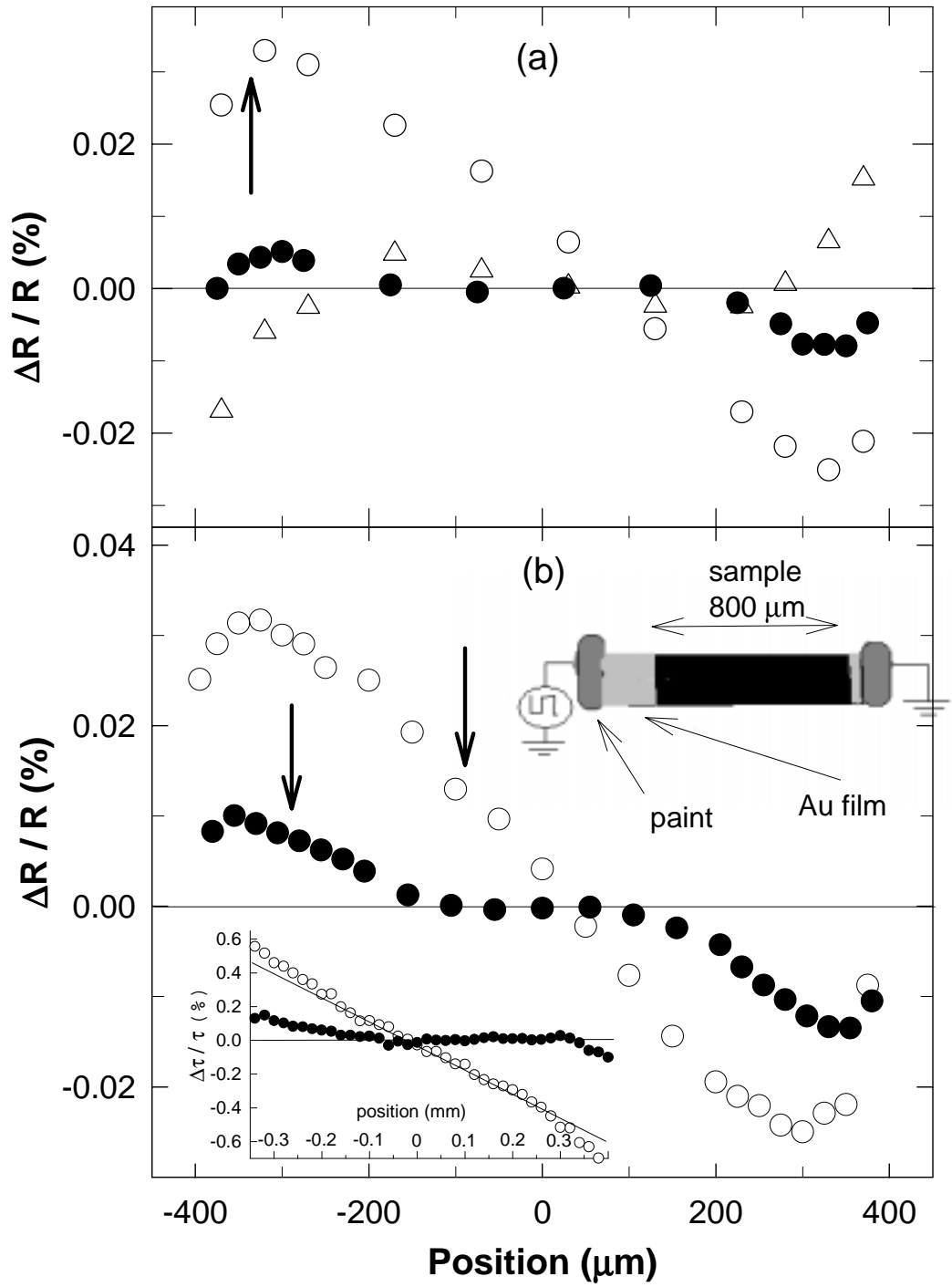

Figure 1

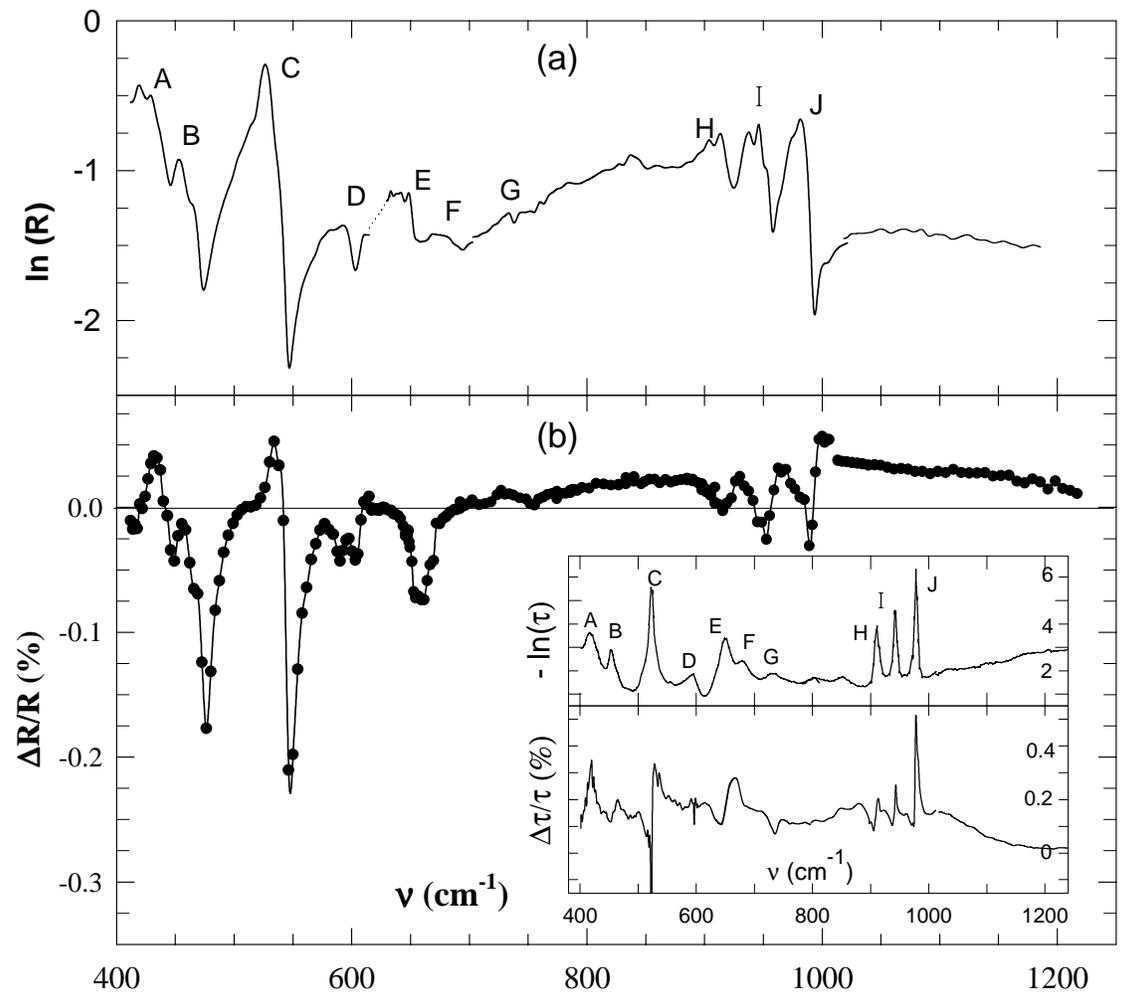

Figure 2

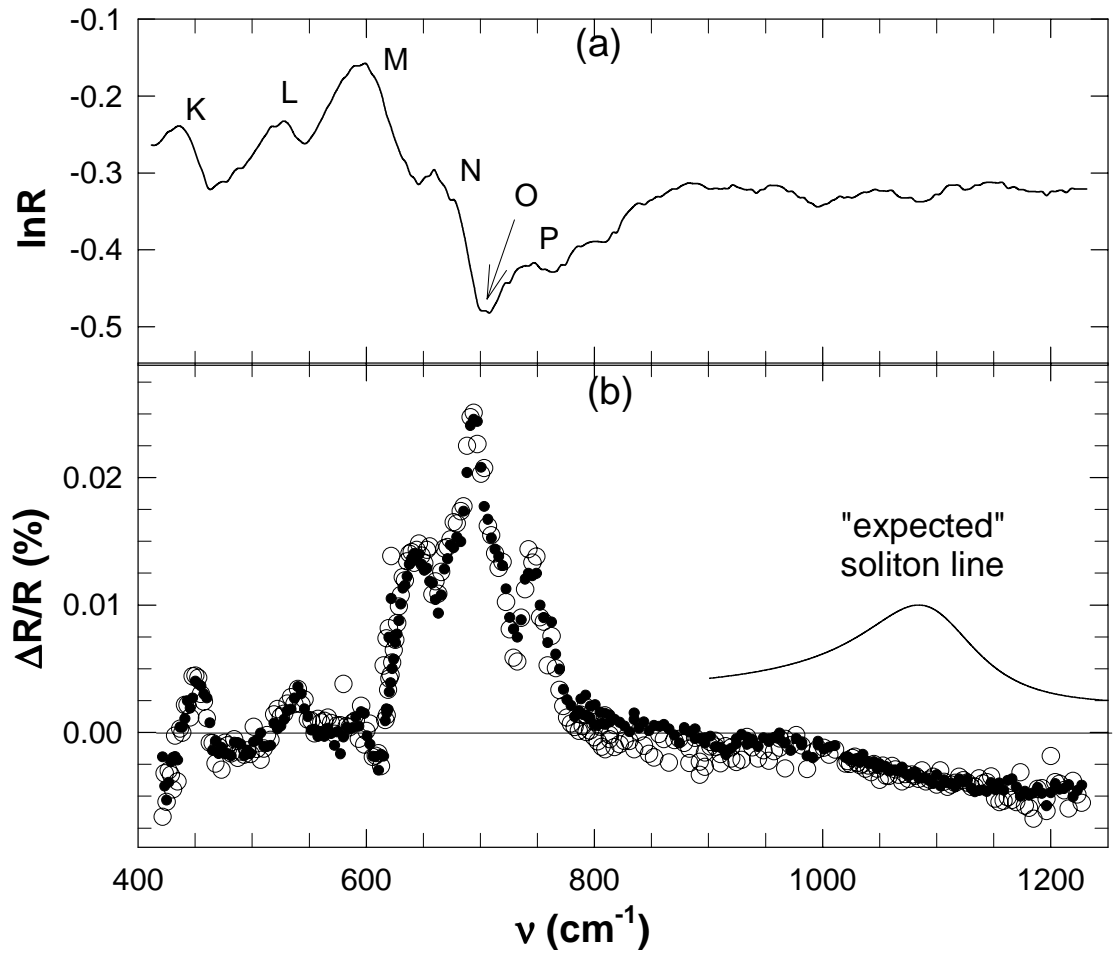

Figure 3